\documentclass[12pt]{article}

\usepackage{setspace}
\usepackage{epsfig}
\usepackage{amssymb}
\def\be{\begin{equation}}
\def\ee{\end{equation}}
\def\ba{\begin{array}}
\def\ea{\end{array}}
\def\beqn{\begin{eqnarray}}
\def\eeqn{\end{eqnarray}}

\def\bt{\begin{tabular}}
\def\et{\end{tabular}}
\def\bc{\begin{center}}
\def\ec{\end{center}}

\begin{document}

\title{Implications of precision measurements on texture specific fermion mass matrices}

\author{Manmohan Gupta$^1$, Gulsheen Ahuja$^1$, Rohit Verma$^2$\\
\\
 {$^1$\it Department of Physics, Centre of Advanced Study, P.U.,
 Chandigarh, India. } \\
{$^2$\it Rayat Institute of Engineering and Information
Technology, Ropar, India. }\\
 {\it Email: mmgupta@pu.ac.in}}

\maketitle

\begin{abstract}
Implications of texture specific mass matrices have been
investigated for both quarks and neutrinos. Interestingly, for the
case of quarks Fritzsch-like texture 4 zero mass matrices have
been found to be compatible with the present precisely known
sin$\,2\beta$ as well as other precise CKM matrix elements. In the
case of leptonic mass matrices, for both Majorana and Dirac
neutrinos we find that for texture 4, 5, 6 zero mass matrices the
inverted hierarchy and degenerate scenarios of neutrino masses are
ruled out by the present data.
\end{abstract}

Understanding fermion masses and mixings constitutes one of the
most important problems of flavor physics. The problem becomes
mysterious when one finds that in the case of quarks the mixing
angles as well as the quark masses show distinct hierarchy, this
being in sharp contrast to the case of neutrinos wherein neither
the mixing angles nor the masses follow a well defined hierarchy.
In particular, at present there is no consensus about neutrino
masses which may show normal/inverted hierarchy or may even be
degenerate. In this context, Fritzsch-like texture specific mass
matrices \cite{quarkstex,neuttex} have given valuable insight in
understanding several features of fermion masses and mixings.
Usually the elements of the quark mass matrices are assumed to be
hierarchical to explain the strongly hierarchical mixing angles,
in contrast to the case of neutrinos where neutrino mixing
parameters do not enforce any hierarchy on the elements of the
mass matrices.

Several parameters in the case of quarks and leptons are now known
with good deal of precision. In the case of quarks, the parameter
sin$\,2\beta$ and the Cabibbo-Kobayashi-Maskawa (CKM) \cite{ckm}
matrix elements $V_{us}, V_{cb}, V_{ud}$ are known precisely
\cite{pdg08}, however, $V_{ub}$ is not that precisely known. In
this context, our recent study \cite{ouruni} indicates that
sin$\,2\beta$ plays a crucial role in fixing $V_{ub}$. Similar to
the case of quarks, in the case of neutrinos angles $s_{12}$,
$s_{23}$ as well as the mass square differences $\Delta
m_{12}^{2}$, $\Delta m_{23}^{2}$ are now rather well known
\cite{pdg08, fogli}, whereas only an upper limit is known
regarding the third mixing angle $s_{13}$. Further, at present not
much is known about the hierarchy of neutrino masses as well as
about their absolute values.

The purpose here, on the one hand, is to check the compatibility
of texture 4 zero mass matrices with the latest precisely known
CKM parameters, in particular with sin$\,2\beta$. On the other
hand, one would like to examine the implications of the present
knowledge of neutrino oscillation parameters on the hierarchy of
neutrino masses in the context of texture specific mass matrices.

Before we examine the compatibility of texture 4 zero mass
matrices with the latest data, we would first like to briefly
discuss the implications of sin$\,2\beta$, along with the
unitarity of the CKM matrix, on the CKM element $V_{ub}$.
Considering the usual `$db$' unitarity triangle and definitions of
the angles of the unitarity triangle, one arrives at a relation
\cite{ouruni} involving CP violating phase $\delta$, the mixing
angles $s_{12}$, $s_{23}$, $s_{13}$ as well as the angle $\beta$
of the unitarity triangle, e.g.,
\be
{\rm tan}\,\frac{\delta}{2} = \frac{A - \sqrt{A^2-(B^2-A^2
C^2){\rm tan}^2 \beta}}{(B+AC){\rm tan}\,\beta}, \label{tand} \ee
where $A=c_{12} s_{12} s_{13}$, $B=c_{23} s_{23}(s_{12}^2-c_{12}^2
s_{13}^2)$ and $C=c_{23}^2-s_{23}^2$. Using $s_{12}^2 \gg c_{12}^2
s_{13}^2$ and $s_{23}^2 \ll
 c_{23}^2$, the above relation can be re-expressed as
\be
\delta~=-\beta+{\rm
sin}^{-1}\left(\frac{s_{12}s_{23}}{c_{12}s_{13}}{\rm
sin}\beta\right). \label{delb}\ee Further, using the closure
property of the angles of the triangle, the mixing angle $s_{13}$
can be written as
\be
s_{13}=\frac{s_{12}s_{23}\,{\rm sin}\,\beta}{c_{12}\,{\rm
sin}\,\alpha}. \label{s13ab} \ee Using these unitarity based
relations we obtain $V_{ub}> 0.00355$, which on using the recently
measured angle $\alpha$ of the unitarity triangle, translates to
$V_{ub}= 0.0035\pm 0.0002$. For our subsequent analysis, we will
be using this value of $V_{ub}$.

After fixing $V_{ub}$, we would now like to examine, in detail,
the compatibility of texture 4 zero mass matrices with the present
value of sin$\,2\beta$ as well as with the other precisely known
CKM elements. To this end, it may be noted that realizing the
importance of the parameter sin$\,2\beta$ for the texture specific
mass matrices, several authors \cite{roberts}-\cite{kimraby} have
explored its implications on these. In particular, using
assumption of strong hierarchy of the elements of the mass matrix,
they arrive at leading order relationships between the various
elements of the mixing matrices and the quark masses. Further,
using these relations and following PDG definition, in the strong
hierarchy case, the angle $\beta$ of the unitarity triangle comes
out to be
\begin{eqnarray} \beta\equiv{\rm arg}\left[-\frac{V_{cd} V_{cb}^*}{V_{td}
 V_{tb}^*}\right]={\rm arg}\left[ 1- \sqrt{\frac{m_u m_s}{m_c m_d}} e^{-i \phi}
  \right].\label{betaothers} \end{eqnarray}

Unfortunately, the value of sin$\,2\beta$ predicted by the above
formula is in quite disagreement with its present precisely known
value. In particular, with the present values of input quark
masses, the value of sin$\,2\beta$ comes out to be $<$ 0.45, which
is in sharp conflict with its present value given by PDG 2008
\cite{pdg08}, e.g., $0.681 \pm 0.025$. This is to be contrasted
with some recent analyses \cite{xingcps, matsuda} which do not
indicate this anomaly. This immediately brings forth the issues of
modification of the above formula as well as of using strongly
hierarchical mass matrices in the context of quarks. In this
context, it may be noted that in principle it is an easy task to
find an exact expression for $\beta$ in the case of texture 4 zero
mass matrices, however such an expression does not yield any
useful information as it involves 4 $V_{{\rm CKM}}$ elements, each
of which has complicated dependence on quark masses, phases and
free parameters of quark mass matrices. Therefore, the first step
in this direction is to develop a formula for sin$\,2\beta$ which
allows one to go beyond the leading order, incorporate the correct
phase structure of the mass matrices as well as allows one to
understand the implications of the hierarchy on the mass matrices.

To facilitate the understanding of the relationship of the present
work with the earlier attempts, we first define the modified
Fritzsch-like matrices, e.g., \be M_i = \left( \ba {ccc} 0 & A_i &
0 \\ A_i^{*} & D_i & B_i
\\
            0 & B_i^{*} & C_i \ea \right), \qquad i=U,D\,, \label{uniq} \ee
$M_U$ and $M_D$, respectively corresponding to the mass matrix in
the up sector and the down sector. It may be noted that each of
the above matrix is texture 2 zero type with $A_{i}
=|A_{i}|e^{i\alpha_{i}}$ and $B_{i} = |B_{i}|e^{i\beta_{i}}$.

Without getting into the details, using only the hierarchy of
quark masses, $m_t \gg m_u$ and $m_b \gg m_d$, one can express
$\beta$ in terms of the quark masses as well as the phases of the
quark mass matrix, e.g.,
 \begin{eqnarray} \beta\equiv{\rm arg}\left[-\frac{V_{cd} V_{cb}^*}{V_{td}
 V_{tb}^*}\right]={\rm arg}\left[ \left( 1- \sqrt{\frac{m_u m_s}{m_c m_d}} e^{-i (\phi_1 + \phi_2)} \right)
 \left( \frac{1-r_2 e^{i \phi_2}}{1-r_1 e^{i \phi_2}} \right)  \right],
 \label{beta} \end{eqnarray} wherein
 \be r_1 = \frac{\zeta_{1 U} \,
\zeta_{3 U}\, \zeta_{1 D}\, \zeta_{2 D}}{\zeta_{2 U}\, \zeta_{3
D}}~~~~~~~~~{\rm and}~~~~~~~~~ r_2 = \frac{\zeta_{1 U} \, \zeta_{3
U}\, \zeta_{2 D}}{\zeta_{2 U}\, \zeta_{1 D}\, \zeta_{3 D}}, \ee
with $\zeta_{1 i}$, $\zeta_{2 i}$, $\zeta_{3 i}$ ($i$ denoting $U$
and $D$) given by
\be
\zeta_{1 i}= \sqrt{1 + \frac{m_2}{C_i}},~~~~\zeta_{2 i}= \sqrt{1 -
\frac{C_i}{m_3}},~~~~\zeta_{3 i}= \sqrt{\frac{C_i}{m_3}}. \ee The
two phases $\phi_1$ and $\phi_2$ are defined as
\be
\phi_1 =  \alpha_U- \alpha_D ~~~~~~~~~~{\rm and} ~~~~~~~~~~
\phi_2= \beta_U- \beta_D.  \ee It may be noted that the
relationship of $\beta$ derived by us, given in equation
(\ref{beta}), is almost an exact formula emanating from texture 4
zero mass matrices, incorporating both the phases $\phi_1$ and
$\phi_2$ as well as no restriction on the hierarchy of the
elements of the mass matrices.

Before discussing the results of our analysis, we would first like
to briefly mention the inputs used for carrying out the
calculations. We have adopted the following ranges of quark masses
\cite{xinginput} at the energy scale of $M_z$, e.g.,
\beqn~~m_u=0.8 -1.8\, {\rm MeV}, ~~~~m_d=1.7 -4.2\, {\rm MeV},
~~~~m_s=40.0 -71.0\, {\rm MeV}, \nonumber\\ m_c=0.63\, {\rm
GeV},~~~~~m_b=2.9\, {\rm GeV},~~~~~m_t=172.5\, {\rm GeV}.
~~~~~~~~~~~~~~~~\eeqn Further, we have given full variation to the
phases $\phi_1$ and $\phi_2$, the parameters $D_U$ and $D_D$ have
respectively been varied from 0 to ($m_t-m_c$) and 0 to
($m_b-m_s$). Furthermore, we have used the constraints imposed by
the following well known CKM parameters \cite{pdg08}, \beqn
|V_{us}|=0.2255\pm 0.0019,~~~~|V_{cb}|=(41.2\pm 1.1) 10^{-3},~~~~
|V_{ub}|= 0.0035\pm 0.0002,\nonumber
\\{\rm sin}\,2\beta = 0.681\pm 0.025.~~~~~~~~~~~~~~~~~~~~~~~~~~~~~~~\eeqn
As mentioned earlier, we have considered the value of $|V_{ub}|$
obtained recently \cite{ouruni} using only the unitarity of the
elements of the CKM matrix and current sin$\,2\beta$ value.

As a first step of our analysis, using these inputs we have
obtained the CKM matrix at 1$\sigma$ C.L. as follows
 \be V_{{\rm CKM}} = \left( \ba{ccc}
  0.9738-0.9747 &~~~~   0.2236-0.2274 &~~~~  0.0033-0.0037 \\
 0.2234-0.2273  &~~~~   0.9729-0.9739    &~~~~  0.0401-0.0423\\
0.0067-0.0103  &~~~~  0.0390-0.0418 &~~~~  0.9991-0.9992 \ea
\right). \label{1sm} \ee A general look at the matrix reveals that
the ranges of CKM elements obtained here are quite compatible with
those obtained by recent global analyses \cite{pdg08, ckmfitter,
utfit, hfag}.

After having checked the compatibility of texture 4 zero mass
matrices with the latest CKM matrix elements, we would now like to
briefly discuss the implications of compatibility of sin$\,2\beta$
on these matrices. Our calculations indicate that sin$\,2\beta$
has important implications for the hierarchy as well as the phase
structure of the mass matrices. It is interesting to note that if
we consider the particular case of strong hierarchy of the
elements of the mass matrices, then even if we use the exact
formula, derived here, given in equation (\ref{beta}), we are not
able to reproduce sin$\,2\beta$ within its experimental limits.
This clearly indicates that hierarchy of mass matrices play an
important role in fitting the data. Further, despite giving full
variation to other parameter and using the exact formula, in case
we consider $\phi_2=0^{\rm o}$, again we are not able to fit
sin$\,2\beta$. These conclusions can be well understood by
comparing the relationship derived by us, given in equation
(\ref{beta}), with the earlier used leading order expression,
given in equation (\ref{betaothers}). Now, the formula has an
additional term $\left( \frac{1-r_2 e^{i \phi_2}}{1-r_1 e^{i
\phi_2}} \right)$ which leads to significant contribution to
sin$\,2\beta$. This happens only in case when $\phi_2 \neq 0^{\rm
o}$ and $D_U/C_U$ or $D_D/C_D$ $>$ 0.07, implying somewhat weak
hierarchy of the elements of the mass matrices.

As a next step of our analysis we have discussed the implications
of the present knowledge of neutrino oscillation parameters on the
texture structure of neutrino mass matrices, examined by us
\cite{ouroldt4, ourmmplb, ourtex4zero}. In particular, we have
studied the issues pertaining to hierarchy of neutrino masses,
e.g., the normal hierarchy case is defined as
$m_{\nu_1}<m_{\nu_2}\ll m_{\nu_3}$, the inverted hierarchy case is
given by $m_{\nu_3} \ll m_{\nu_1} < m_{\nu_2}$ and the
corresponding degenerate cases are respectively defined as
$m_{\nu_1} \lesssim m_{\nu_2} \sim m_{\nu_3}$ and $m_{\nu_3} \sim
m_{\nu_1} \lesssim m_{\nu_2}$. For both Majorana as well as Dirac
neutrinos, we have imposed Fritzsch-like texture structure on
Dirac neutrino mass matrix, with charged leptons having either
Fritzsch-like texture structure or being in the flavor basis.

Considering the modified Fritzsch-like matrices, e.g.,
 \be
 M_{l}=\left( \ba{ccc}
0 & A _{l} & 0      \\ A_{l}^{*} & D_{l} &  B_{l}     \\
 0 &     B_{l}^{*}  &  C_{l} \ea \right), \qquad
M_{\nu D}=\left( \ba{ccc} 0 &A _{\nu} & 0      \\ A_{\nu}^{*} &
D_{\nu} &  B_{\nu}     \\
 0 &     B_{\nu}^{*}  &  C_{\nu} \ea \right),
 \label{frzmm7}
 \ee
$M_{l}$ and $M_{\nu D}$ respectively corresponding to Dirac-like
charged lepton and neutrino mass matrices, we have investigated 42
distinct possibilities of texture 4 zero, 5 zero and 6 zero mass
matrices for normal/inverted hierarchy as well as degenerate
scenarios of neutrino masses. The texture 6 zero matrices
correspond to the above mentioned matrices when both $D_l$ and
$D_{\nu}$ are zero whereas texture 5 zero matrices correspond to
either $D_l=0$ and $D_{\nu}\neq 0$ or $D_{\nu}=0$ and $D_l \neq
0$, the above matrices are texture 4 zero when both $D_l$ and
$D_{\nu}$ are non-zero.

The latest situation regarding masses and mixing angles at
3$\sigma$ C.L. is summarized as follows \cite{fogli},
\be
 \Delta m_{12}^{2} = (7.14 - 8.19)\times
 10^{-5}~\rm{eV}^{2},~~~~
 \Delta m_{23}^{2} = (2.06 - 2.81)\times 10^{-3}~ \rm{eV}^{2},
 \label{solatmmass}\ee
\be
{\rm sin}^2\,\theta_{12}  =  0.263 - 0.375,~~~
 {\rm sin}^2\,\theta_{23}  =  0.331 - 0.644,~~~
 {\rm sin}^2\,\theta_{13} \leq 0.046. \label{s13}
\ee

 To begin with, for Fritzsch-like texture 4, 5 and 6 zero
lepton mass matrices detailed predictions for cases pertaining to
inverted hierarchy as well as degenerate scenarios of neutrino
masses have been carried out. For Majorana as well as Dirac
neutrinos, our rigorous analysis, wherein all the input parameters
have been given 3$\sigma$ variations, reveals that all these cases
seem to be ruled out. For the present purpose, we illustrate the
case of texture 4 zero mass matrices. In figures 1a, 1b and 1c,
for Majorana neutrinos we have plotted the parameter space
corresponding to any of the two mixing angles by constraining the
third angle by its values given in equation (\ref{s13}) while
giving full allowed variation to other parameters. Also included
in the figures are blank rectangular regions indicating the
experimentally allowed $3\sigma$ region of the plotted angles.
Interestingly, a general look at these figures reveals that the
case of inverted hierarchy seems to be ruled out. From figure 1a
showing the plot of angles $\theta_{12}$ versus $\theta_{23}$, one
can immediately conclude that the plotted parameter space includes
the experimentally allowed range of
$\theta_{23}=35.7^{\circ}-55.6^{\circ}$, however it excludes the
experimentally allowed range of
$\theta_{12}=29.3^{\circ}-39.25^{\circ}$. This clearly indicates
that at 3$\sigma$ C.L. inverted hierarchy is not viable. The
conclusions arrived above can be further checked from figures 1b
and 1c wherein we have plotted $\theta_{12}$ versus $\theta_{13}$
and $\theta_{23}$ versus $\theta_{13}$ respectively by
constraining angles $\theta_{23}$ and $\theta_{12}$. Without
getting into the details, one can show that the degenerate
scenarios mentioned above are also ruled out for these mass
matrices. Further, from the above analysis of texture 4 zero mass
matrices, one can easily deduce similar conclusions for texture 6
zero as well as for the two cases of the texture 5 zero mass
matrices.

\begin{figure}
\psfig{file=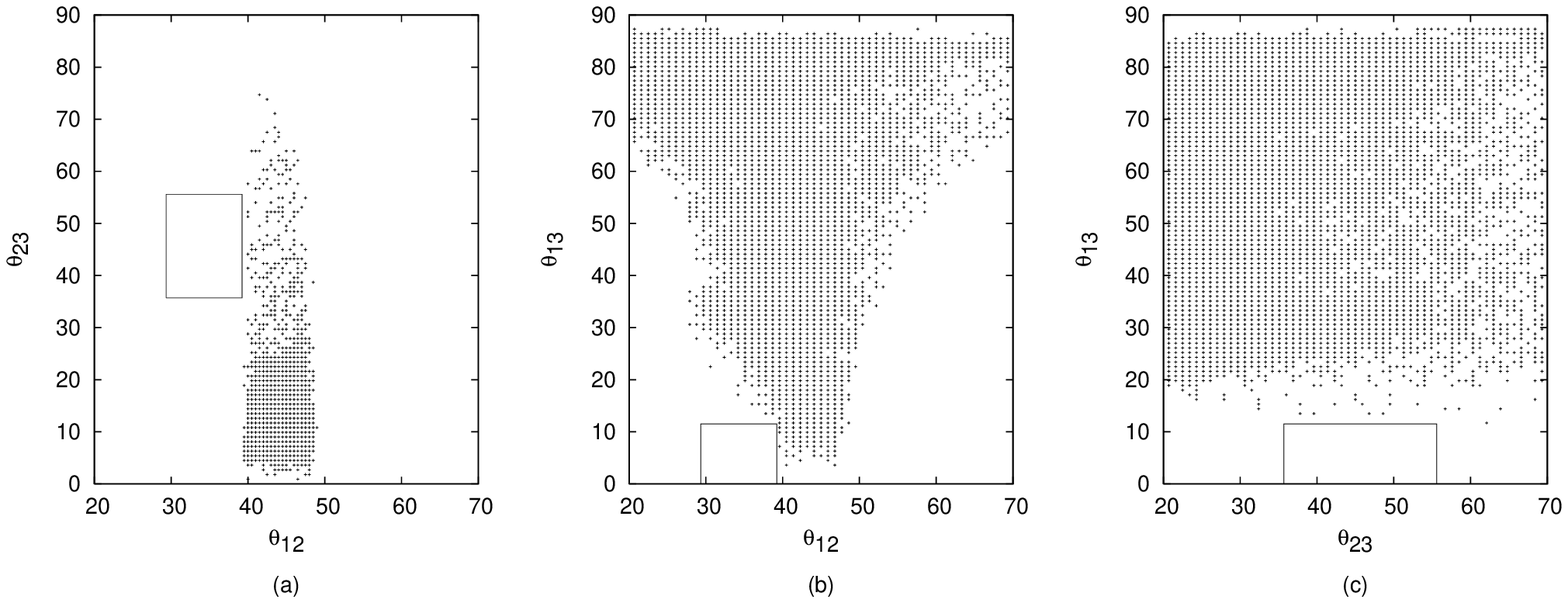,width=5.in} \caption{Plots showing
the parameter space corresponding to any of the two mixing angles
by constraining the third angle by its experimental limits given
in equation (\ref{s13}) and giving full allowed variation to other
parameters for Majorana neutrinos. The blank rectangular region
indicates the experimentally allowed $3\sigma$ region of the
plotted angles.}
  \label{fig1}
  \end{figure}

Coming to the normal hierarchy cases, detailed dependence of
mixing angles on the lightest neutrino mass as well as the
parameter space available to the phases of mass matrices have been
investigated for both Majorana and Dirac neutrinos \cite{ourmmplb,
ourtex4zero}. Without presenting the details, several
phenomenological quantities such as the viable ranges of neutrino
masses, mixing angle $s_{13}$, Jarlskog's rephasing invariant
parameter $J_l$, the Dirac-like CP violating phase $\delta_l$ and
the effective neutrino mass $ \langle m_{ee} \rangle$, related to
neutrinoless double beta decay $(\beta\beta)_{0 \, \nu}$, have
also been calculated for different cases. In the case of texture 4
zero mass matrices, for Majorana neutrinos, we obtain
$m_{\nu_1}=2.47 \times 10^{-4} - 0.006$, $\theta_{13}=1.14^{\circ}
- 11.50^{\circ}$ and $J_l=-0.0459 - .0463$, whereas for Dirac
neutrinos, these come out to be $m_{\nu_1}=5.73 \times 10^{-5}  -
0.012$, $\theta_{13}=0.084^{\circ} - 11.50^{\circ}$ and
$J_l=-0.0462 - .0448$. Our analysis reveals that a measurement of
$m_{\nu_1}$ could have important implications for the nature of
neutrinos. Also, the lower limit on $\theta_{13}$ for the Dirac
case is considerably lower than for the Majorana case, therefore a
measurement of $\theta_{13}$ would have important implications for
this case.

For the sake of completion, we have also carried out calculations
wherein the charged lepton mass matrix is in the flavor basis for
all the texture specific cases considered here. As expected, the
inverted hierarchy and degenerate scenarios are ruled out for all
the Dirac and Majorana cases. For the normal hierarchy, the
texture 6 zero and the texture 5 zero $D_{\nu}=0$ case are again
ruled out. For the texture 5 zero $D_l=0$ case and the texture 4
zero mass matrices, the ranges of various phenomenological
quantities calculated here are much narrower compared to the
corresponding texture specific cases.

To summarize, implications of Fritzsch-like Hermitian texture
specific mass matrices have been investigated for both quarks and
neutrinos. In the case of quarks, our analysis reveals that
texture 4 zero mass matrices are quite compatible with well known
parameter sin$\,2\beta$ and with other precise CKM matrix
elements. In the case of leptonic mass matrices, for Majorana and
Dirac neutrinos, we find that for texture 4, 5, 6 zero mass
matrices the inverted hierarchy and degenerate scenarios of
neutrino masses are ruled out by the present data.

\vskip 0.5cm {\bf Acknowledgements} \\ GA would like to thank DST,
Government of India for financial support and the Chairman,
Department of Physics for providing facilities to work. RV would
like to thank the Director, RIEIT, for providing facilities to
work.

\end{document}